\documentclass[twocolumn,american]{article}
\usepackage[T1]{fontenc}
\usepackage[latin9]{inputenc}
\usepackage{babel}
\usepackage{amsmath}
\usepackage{amssymb}
\usepackage{graphicx}
\usepackage{geometry}
\usepackage[pdfusetitle,
 bookmarks=true,bookmarksnumbered=false,bookmarksopen=false,
 breaklinks=false,pdfborder={0 0 1},backref=false,colorlinks=false]
 {hyperref}

\makeatletter
\@ifundefined{date}{}{\date{}}

\makeatother

\begin{document}
\title{The electromagnetic stress tensor in cubic crystals and amorphous
solids}
\author{Richard Dengler \thanks{ORCID: 0000-0001-6706-8550}\\
Munich, Germany}
\maketitle
\begin{abstract}
The electromagnetic vacuum stress tensor in a system consisting of
classical point-like atoms on a cubic lattice is derived directly
from the electric field and, alternatively, using the known more phenomenological
formula involving the (usually unknown) derivative of the dielectric
tensor. The results agree. The stress tensor contains a nontrivial
constant and therefore cannot be obtained from macroscopic electrodynamics
alone.

A system consisting of a mosaic of randomly oriented grains of a cubic
crystal is used as a model of an amorphous solid. The electromagnetic
stress tensor in this system is deduced by averaging the corresponding
quantity in a cubic crystal over all orientations. Two components
remain: the component in the direction of the electric field and the
component perpendicular to it. The transverse component agrees with
the analogous quantity for liquids derived by Peierls in an entirely
different way.
\end{abstract}

\section{Introduction}

Macroscopic electrodynamics is an astonishingly successful reformulation
of the microscopic Maxwell equations, applicable when the characteristic
length scales of the system in question are much larger than the atomic
scale. It correctly describes quasi-static electromagnetic phenomena
in condensed matter, as well as the reflection and refraction of light
by dielectric media for different polarizations and the intensities
of reflected and transmitted light in the isotropic and anisotropic
case~\cite{Lipson1981,Jackson1999}. Material properties enter solely
through constitutive coefficients such as susceptibilities and electrical
conductivity. 

The fact that the theory, which is essentially linear in the fields,
also correctly reproduces the bilinear energy density, is an additional
bonus. The underlying physical principle is rarely stated. Atomic
polarization currents are additive and are equivalent to a time-dependent
macroscopic surface charge at the boundary of a medium, for instance
on the plates of a capacitor. Similarly, the microscopic circular
currents in the atoms of a magnetizable cylinder cancel in the interior
but are equivalent to a macroscopic current around the cylinder. The
macroscopic quantities then determine the energy. 

Difficulties arise, however, for another bilinear quantity, the electromagnetic
stress tensor, or momentum flux. This limitation is the origin of
longstanding debates. Two more recent reviews are \cite{Milonni_2010,Kemp2011}.
Rather than addressing the problem in a purely formal manner, we
here calculate the stress tensor exactly for microscopic models. Such
models have been used for various purposes~\cite{Draine_1993,Vries_1998,Draine_1988,LoWanYu2001}.
The microscopic stress tensor, however, has not been previously evaluated.

\section{Electromagnetic stress tensor in a lattice of point dipoles}

\label{sec:sigma_cubic_E1}The central tool used in this work is the
Fourier series of the electric field in an orthorhombic lattice of
point dipoles. We derive this sum from a model of small, but finite,
classical polarizable atoms.

The result reproduces the electric field of the discrete-dipole approximation~\cite{Draine_1988,Draine_1993}
if the atoms are small, and also coincides with the Ewald-Kornfeld
sum~\cite{Kornfeld1924,LoWanYu2001} if the convergence parameter
of this formalism is chosen in such a way the the reciprocal lattice
sum dominates.

A direct result is a formula expressing the components $\epsilon_{ii}$
of the dielectric tensor in terms of the polarizability of an atom,
which reduces to the Clausius-Mossotti relation in the case of cubic
crystals, but also yields the dependence of $\epsilon_{ii}$ on deformations
of the crystal. This dependence is needed to compute the stress tensor
from the known phenomenological formula.

\subsection{The electric field}

The atoms are assumed to consist of two interpenetrating spherically
symmetric charge distributions $\pm Q_{a}F\left(\boldsymbol{x}\right)$
with a charge $\pm Q_{a}$. We assume that the radial scale $R$ of
$F\left(\boldsymbol{x}\right)$ satisfies $R\ll\min\left(a_{1},a_{2},a_{3}\right)$,
where $a_{i}$ are the lattice constants of the orthorhombic crystal.
Three simple charge distributions and their Fourier transform $\tilde{F}$
are
\begin{align}
\begin{array}{c|c}
F\left(\boldsymbol{x}\right) & \tilde{F}\left(qR\equiv s\right)\\
\hline e^{-\boldsymbol{x^{2}}/R^{2}}/\left(\sqrt{\pi}R\right)^{3} & e^{-s^{2}/4}\\
e^{-\left|\boldsymbol{x}\right|/R}/\left(8\pi R^{3}\right) & 1/\left(1+s^{2}\right)^{2}\\
\theta\left(R^{2}-\boldsymbol{x}^{2}\right)/\left(\tfrac{4\pi}{3}R^{3}\right) & 3\left(\sin s-s\cos s\right)/s^{3}.
\end{array}\label{eq:F_Funct}
\end{align}
We assume that $Q_{a}$ is large. The displacements of the charges
then are small, and the charge density of the atom at the origin in
a local field $\boldsymbol{E}^{\mathrm{loc}}$ is
\begin{align}
\rho^{\#\left(0\right)}_{\mathrm{dip}}\left(\boldsymbol{x}\right) & =-\boldsymbol{p}\nabla F\left(\boldsymbol{x}\right)=-\epsilon_{0}\gamma V\boldsymbol{E}^{\mathrm{loc}}\cdot\nabla F\left(\boldsymbol{x}\right),\label{eq:rho_dip0}
\end{align}
where
\begin{equation}
\boldsymbol{p}=\epsilon_{0}\gamma V\boldsymbol{E}^{\mathrm{loc}}\label{eq:gamma_def}
\end{equation}
is the polarization of an atom, and $V=a_{1}a_{2}a_{3}$ the unit
cell volume. The symbol '\#' always denotes exact, microscopic, quantities.
The polarizability of an atom is $\gamma_{a}=\gamma V$. The charge
density of the crystal is periodic and can be written as a Fourier
series,
\begin{align*}
\rho^{\#}\left(\boldsymbol{x}\right) & =-i\epsilon_{0}\gamma\boldsymbol{E}^{\mathrm{loc}}\cdot\sum_{q_{i}\in2\pi\mathbb{Z}/a_{i}\setminus0}e^{i\boldsymbol{qx}}\boldsymbol{q}\tilde{F}\left(qR\right),\\
\tilde{F}\left(qR\right) & =\int\mathrm{d}^{3}xe^{-i\boldsymbol{qx}}F\left(\boldsymbol{x}\right).
\end{align*}
We now assume that $\boldsymbol{E}^{\mathrm{loc}}$ and the average
(macroscopic) field $\boldsymbol{E}$ point in $x_{1}$ direction.
The Fourier series of the electric field then is
\begin{align}
\boldsymbol{E}^{\#}\left(\boldsymbol{x}\right) & =\boldsymbol{E}+\tilde{\boldsymbol{E}}\left(\boldsymbol{x}\right),\label{eq:E=000023_ortho}\\
\tilde{\boldsymbol{E}}\left(\boldsymbol{x}\right) & =-\gamma E^{\mathrm{loc}}_{1}\sum_{\boldsymbol{q}\in\left(2\pi\mathbb{Z}/a_{i}\right)\setminus0}e^{i\boldsymbol{qx}}\tfrac{\boldsymbol{q}q_{1}}{q^{2}}\tilde{F}\left(qR\right)\cdot\nonumber 
\end{align}
It suffices to verify that the Maxwell equations $\nabla\cdot\boldsymbol{\tilde{\boldsymbol{E}}}=\rho^{\#}/\epsilon_{0}$
and $\nabla\times\boldsymbol{\tilde{\boldsymbol{E}}}=0$ are satisfied.
The usual definition $p_{1}/V=\left(\epsilon_{11}-\epsilon_{0}\right)E_{1}$
of the dielectric constant in terms of polarization density $p_{1}$
and average field $E_{1}$ leads to
\begin{equation}
\left(\epsilon_{11}-\epsilon_{0}\right)E_{1}=\epsilon_{0}\gamma E^{\mathrm{loc}}_{1}.\label{eq:eps_1_gamma}
\end{equation}
This equation can be used to replace the microscopic quantity $\gamma E^{\mathrm{loc}}_{1}$
in the expression~(\ref{eq:E=000023_ortho}) for the electric field
with a macroscopic quantity. The computation of the stress tensor
then is a purely formal task.

However, to relate the stress tensor to the conventional expression
we also need the dielectric constant, which can be derived by writing
the field $\boldsymbol{E}^{\#}\left(\boldsymbol{x}=0\right)$ at the
origin in a different way. The alternative expression is the sum of
$E^{\mathrm{loc}}_{1}$ and the field generated by the charge distribution~(\ref{eq:rho_dip0})
of the atom at the origin at $\boldsymbol{x}=0$,
\begin{equation}
E^{\mathrm{dip}}_{i}=\tfrac{-1}{4\pi\epsilon_{0}}\int\mathrm{d}^{3}x\rho^{\#\left(0\right)}_{\mathrm{dip}}\left(\boldsymbol{x}\right)\nabla_{i}\left|\boldsymbol{x}\right|^{-1}=-\tfrac{\gamma}{3}VF\left(0\right)E^{\mathrm{loc}}_{1}\delta_{i,1}.\label{eq:E=0000230_E_loc}
\end{equation}
Equating now $E^{\mathrm{loc}}_{1}+E^{\mathrm{dip}}_{i}$ and the
expression (\ref{eq:E=000023_ortho}) for $\boldsymbol{x}=0$ and
eliminating $E_{1}$ with (\ref{eq:eps_1_gamma}) leads to
\begin{align}
\epsilon_{11}/\epsilon_{0} & =1+\gamma/\left(1-\gamma\Lambda\right),\label{eq:eps1_from_sum}\\
\Lambda & =\tfrac{V}{3}F\left(0\right)-\sum_{\boldsymbol{q}\in\left(2\pi\mathbb{Z}/a_{i}\right)\setminus0}\tfrac{q^{2}_{1}}{q^{2}}\tilde{F}\left(qR\right).\label{eq:Lambda_def}
\end{align}
Equations (\ref{eq:eps1_from_sum}) and (\ref{eq:Lambda_def}) determine
the dielectric constant of orthorhombic crystals. Atoms with a Gaussian
charge distribution $\tilde{F}\left(qR\right)=e^{-q^{2}R^{2}/4}$
yield a rapidly convergent sum. In fact, with this choice Eq.~(\ref{eq:Lambda_def})
agrees with the reciprocal lattice component of the corresponding
Ewald-Kornfeld sum for point dipoles, the radius $R$ plays the role
of the convergence parameter. The lattice component is of the order
$e^{-a^{2}/R^{2}}$ and negligible for $R\ll a$. This confirms that
the limit $R\rightarrow0$ is equivalent to point dipoles. Consequently,
full Ewald summation is often unnecessary with modern computing power;
the reciprocal-lattice sum converges quickly, and the real-lattice
contribution remains negligible for sufficiently small $R$.

For cubic crystals and for $R\ll a$ the sum in (\ref{eq:Lambda_def})
can be approximated by an integral if the missing $q=0$ term $1/3$
is added. This gives $\sum_{q\neq0}\ldots=-1/3+VF\left(0\right)/3$
and $\Lambda=1/3$. This value together with (\ref{eq:eps1_from_sum})
reproduces the usual Clausius-Mossotti equation for $\epsilon$ in
a cubic crystal. 

\subsection{Stress tensor perpendicular to the electric field}

We now use the microscopic electric field (\ref{eq:E=000023_ortho})
to compute the electromagnetic vacuum stress tensor
\begin{align}
\sigma^{\#}_{ij}\left(\boldsymbol{E}^{\#},\boldsymbol{B}^{\#}\right) & =\tfrac{\epsilon_{0}}{2}\left[\delta_{ij}\left(\boldsymbol{E}^{2}_{\#}+\boldsymbol{B}^{2}_{\#}\right)-2E^{\#}_{i}E^{\#}_{j}-2B^{\#}_{i}B^{\#}_{j}\right]\label{eq:sigma=000023def}
\end{align}
Our sign convention for $\sigma^{\#}_{ij}$ is that of \cite{LL_FT1971}:
a positive diagonal component compresses the medium in the given direction.
We begin with the interstitial vacuum plane $A_{3}$ defined by $x_{3}=a/2$,
parallel to the field $E_{1}$. The computation in the plane $A_{2}$
is analogous.

The granular part (\ref{eq:E=000023_ortho}) of $\boldsymbol{E}^{\#}$
can be written as
\begin{align}
\tilde{E}_{i} & =\left(\epsilon_{r}-1\right)E_{1}\partial_{i}\partial_{1}\tilde{S},\label{eq:E_tilde}\\
\tilde{S} & =\sum_{\boldsymbol{q}\in\left(2\pi\mathbb{Z}/a\right)^{3},q_{1}\neq0}e^{i\boldsymbol{qx}}/\boldsymbol{q}^{2}.\label{eq:S_sum}
\end{align}
The derivative $\partial_{1}$ in $\tilde{E}$ allows to restrict
the sum in $\tilde{S}$ to $q_{1}\neq0$ instead of merely $\boldsymbol{q}\neq0$.
The sum over $q_{3}$ in $\tilde{S}$ can now be performed using formula
(\ref{eq:MatsubaraSum}) from Appendix~\ref{subsec:App_FT_sum}.
The result is
\begin{align}
\tilde{S} & =\tfrac{a^{2}}{4}\sum_{q_{1}\neq0,q_{2}}e^{i\left(q_{1}x_{1}+q_{2}x_{2}\right)}\tfrac{\cosh\left(Q\left(x_{3}-\tfrac{a}{2}\right)\right)}{\tfrac{Qa}{2}\sinh\tfrac{Qa}{2}},\label{eq:STildeSum}
\end{align}
with $Q^{2}=q^{2}_{1}+q^{2}_{2}$. For $x_{3}=a/2$ this leads to
\begin{equation}
\tilde{E}_{i\in\left\{ 1,2\right\} }\left(A_{3}\right)=-\left(\epsilon_{r}-1\right)E_{1}\sum_{q_{1}\neq0,q_{2}}e^{i\left(q_{1}x_{1}+q_{2}x_{2}\right)}\tfrac{a^{2}q_{i}q_{1}/4}{\tfrac{Qa}{2}\sinh\tfrac{Qa}{2}}.\label{eq:ETildeSides}
\end{equation}
The field component $\tilde{E}_{3}\left(A_{3}\right)$ vanishes. The
stress tensor component perpendicular the field $E_{1}$ is $\sigma^{\#}_{33}=\tfrac{\epsilon_{0}}{2}\left(E^{\#2}_{1}+E^{\#2}_{2}\right)$.
We are only interested in the average $\left\langle \sigma^{\#}_{33}\right\rangle $
over the area $0\leq x_{1},x_{2}\leq a$ in $A_{3}$. The average
$\left\langle \tilde{E}_{i}\tilde{E}_{j}\right\rangle $ of a product
of two Fourier series like Eq.~(\ref{eq:ETildeSides}) is the sum
of the product of the Fourier components, see Appendix \ref{subsec:App_FT_avg}.
The result is 
\begin{align}
\left\langle \tilde{\sigma}_{33}\left(A_{3}\right)\right\rangle  & =\tfrac{\epsilon_{0}}{2}\left(\epsilon_{r}-1\right)^{2}E^{2}_{1}M,\nonumber \\
M & =\tfrac{1}{2}\sum_{\left\{ q_{1},q_{2}\right\} \neq0}\tfrac{\left(Qa/2\right)^{2}}{\sinh^{2}\tfrac{Qa}{2}}=\text{0.171599}\ldots,\label{eq:M_Def}\\
\sigma^{\mathrm{cubic}}_{33}\left(E_{1}\right) & =\tfrac{\epsilon_{0}}{2}E^{2}_{1}\left(1+\left(\epsilon_{r}-1\right)^{2}M\right),\label{eq:sigma=00002333}
\end{align}
where $M$ is a constant. We have omitted the area $A_{3}$ in the
final expression, since it follows from the indices, and because $\sigma^{\mathrm{cubic}}_{33}$
is the same in any vacuum plane parallel to $A_{3}$. 

\subsection{Stress tensor in direction of the electric field}

To get the electric field in the interstitial planes $A_{1}$ perpendicular
to the field $E_{1}$ we now perform the sum over $q_{1}$ in Eq.~(\ref{eq:S_sum}).
We split the sum into contributions with $Q^{2}=q^{2}_{2}+q^{2}_{3}=0$
and $Q^{2}\neq0$,
\begin{align}
\tilde{S} & =\sum_{q_{1}\neq0}e^{iq_{1}x_{1}}/q^{2}_{1}+\sum_{\boldsymbol{q}\in\left(2\pi\mathbb{Z}/a\right)^{3},\,Q\neq0,\,q_{1}\neq0}e^{i\boldsymbol{qx}}/\boldsymbol{q}^{2}\label{eq:STilde_q2q3}\\
 & =\tfrac{1}{2}\left(x^{2}_{1}-x_{1}a+\tfrac{a^{2}}{6}\right)+\tfrac{a^{2}}{4}\sum_{\left\{ q_{2},q_{3}\right\} \neq0}e^{i\left(q_{2}x_{2}+q_{3}x_{3}\right)}\tfrac{\cosh\left(Q\left(x_{1}-\tfrac{a}{2}\right)\right)}{\tfrac{Qa}{2}\sinh\tfrac{Qa}{2}}.\nonumber 
\end{align}
The condition $q_{1}\neq0$ in the $Q\neq0$ part can be ignored,
as we are only interested in contributions that depend on $x_{1}$.
The sums over $q_{1}$ in Eq.~(\ref{eq:STilde_q2q3}) were performed
using the formulas~(\ref{eq:MatsubaraSum2}) and (\ref{eq:MatsubaraSum})
in Appendix \ref{subsec:App_FT_sum}. From $\partial_{1}\tilde{S}=0$
at $x_{1}=a/2$ it follows that $\tilde{E}_{2}=\tilde{E}_{3}=0$,
but there remains
\begin{align}
\tilde{E}_{1}\left(A_{1}\right) & =\left(\epsilon_{r}-1\right)E_{1}\partial^{2}_{1}\tilde{S}\label{eq:ETildeTop}\\
 & =\left(\epsilon_{r}-1\right)E_{1}\left(\underline{1}+\sum_{\left\{ q_{2},q_{3}\right\} \neq0}e^{i\left(q_{2}x_{2}+q_{3}x_{3}\right)}\tfrac{Qa/2}{\sinh\tfrac{Qa}{2}}\right).\nonumber 
\end{align}
The constant in Eq.~(\ref{eq:ETildeTop}) implies that the actual
non-granular field in $A_{1}$ is $\epsilon_{r}E_{1}$ instead of
$E_{1}$. Averaging $\sigma^{\#}_{11}=-\tfrac{\epsilon_{0}}{2}E^{\#2}_{1}$
over an area $a^{2}$ yields (see Appendix \ref{subsec:App_FT_avg})
\begin{equation}
\sigma^{\mathrm{cubic}}_{11}\left(E_{1}\right)=-\tfrac{\epsilon_{0}}{2}E^{2}_{1}\left(\epsilon^{2}_{r}+\left(\epsilon_{r}-1\right)^{2}2M\right).\label{eq:sigma=00002311}
\end{equation}
Expressions (\ref{eq:sigma=00002333}) and (\ref{eq:sigma=00002311})
for cubic crystals are the main result of this section. The constant
$M$ indicates that the electromagnetic stress tensor, in general,
cannot be derived from macroscopic electrodynamics alone.

\section{Electromagnetic stress tensor from phenomenological formula}

In textbooks one finds the expression~\cite{Jackson1999,LL_FT1971}
\begin{align}
\sigma^{\mathrm{liq}}_{ij}\left(\boldsymbol{E},\boldsymbol{B}\right) & =\tfrac{\epsilon_{0}}{2}\left[\delta_{ij}\left(\left(\epsilon_{r}-\rho\tfrac{\partial\epsilon_{r}}{\partial\rho}\right)\boldsymbol{E}^{2}+\boldsymbol{B}^{2}\right)\right.\label{eq:sigma_LL}\\
 & \qquad\left.-2\epsilon_{r}E_{i}E_{j}-2B_{i}B_{j}\right]\nonumber 
\end{align}
for the electromagnetic stress tensor in non-magnetizable liquids.
The symbols $\boldsymbol{E}$ and $\boldsymbol{B}$ denote the macroscopic
(average) fields, $\epsilon_{r}=\epsilon/\epsilon_{0}$ is the relative
dielectric constant of the liquid, and $\rho$ its mass density. The
derivative $\partial\epsilon_{r}/\partial\rho$ is to be taken at
constant temperature. This expression follows by equating the change
of the electromagnetic energy in a volume to the work done when the
volume is changed~\cite{LL_FT1971}. The practical value of formula
(\ref{eq:sigma_LL}) is limited, however, because the derivative $\partial\epsilon_{r}/\partial\rho$
is usually unknown. In general it also seems difficult to identify
(\ref{eq:sigma_LL}) with some average of the exact microscopic $\sigma^{\#}_{ij}$.
An average over a plane in the liquid would contain contributions
from electromagnetic forces within the atoms, which are compensated
by some other forces, but differ from zero also for $\boldsymbol{E}=0$.

The phenomenological derivation of the stress tensor (\ref{eq:sigma_LL})
can straightforwardly be generalized to isotropic solids and to orthorhombic
crystals if the electric field points in the direction of a crystal
axis, see Appendix~\ref{subsec:App_EnergyBalance}. We here again
use the $x_{1}$ axis. We only consider amorphous media or cubic crystals,
originally of volume $\ell_{1}\ell_{2}\ell_{3}$ and with $\epsilon_{ij}=\delta_{ij}\epsilon$.
An uniaxially compressed liquid remains isotropic and thus $\epsilon=\epsilon\left(\rho\right)$.
This is not true for solids, where a uniaxial deformation changes
the aspect ratio of the unit cells. In general thus $\ell_{1}\partial\epsilon_{11}/\partial\ell_{1}\neq\ell_{3}\partial\epsilon_{11}/\partial\ell_{3}$.
The energy balance leads to
\begin{align}
\sigma^{\mathrm{solid}}_{11}\left(E_{1}\right) & =-\tfrac{1}{2}\left(\epsilon_{11}-\ell_{1}\tfrac{\partial\epsilon_{11}}{\partial\ell_{1}}\right)E^{2}_{1},\label{eq:sigma11_solid}\\
\sigma^{\mathrm{solid}}_{33}\left(E_{1}\right) & =\tfrac{1}{2}\left(\epsilon_{11}+\ell_{3}\tfrac{\partial\epsilon_{11}}{\partial\ell_{3}}\right)E^{2}_{1}.\label{eq:sigma33_solid}
\end{align}
We now compare the phenomenological expressions~(\ref{eq:sigma11_solid})
and (\ref{eq:sigma33_solid}) with the exact results derived above.

\subsection{Cubic crystal}

To actually get a value from the expressions (\ref{eq:sigma11_solid})
and (\ref{eq:sigma33_solid}) we also need the variation of $\epsilon_{11}$
of a cubic crystal under deformations $a\rightarrow a\xi_{i}$ with
$\xi_{i}\cong1$. This variation can be  accounted for in the sum
(\ref{eq:Lambda_def}) by replacing $q_{i}$ with $q_{i}/\xi_{i}$.
The numerical evaluation of (\ref{eq:Lambda_def}) for small deformations
in the direction $\hat{\boldsymbol{x}}_{1}$ or $\hat{\boldsymbol{x}}_{3}$
leads to
\begin{equation}
\epsilon_{r}=1+\tfrac{\gamma_{a}}{V-\gamma_{a}\left[1/3-2\left(M+\tfrac{1}{3}\right)\delta\xi_{1}+\left(M+\tfrac{1}{3}\right)\delta\xi_{3}\right]}+\left(\delta\xi\right)^{2}\ldots,\label{eq:eps_r_ortho}
\end{equation}
where $V=a^{3}\xi_{1}\xi_{2}\xi_{3}$ and $M\cong0.171599$ is the
constant from Eq.~(\ref{eq:sigma=00002333}).

We first consider a deformation in the direction perpendicular to
the electric field. The derivative follows as
\begin{equation}
\left.\partial_{\xi_{3}}\epsilon_{r}\right|_{\xi=1}=-\tfrac{\gamma_{a}\left(a^{3}-\left(M+\tfrac{1}{3}\right)\gamma_{a}\right)}{\left(a^{3}-\gamma_{a}/3\right)^{2}}=-\left(\epsilon_{r}-1-M\left(\epsilon_{r}-1\right)^{2}\right).\label{eq:Deriv3}
\end{equation}
The value is negative for $\epsilon_{r}<1+1/M\cong6.83.$  Inserting
the derivative (\ref{eq:Deriv3}) into the formula (\ref{eq:sigma33_solid})
reproduces the exact average vacuum stress tensor component (\ref{eq:sigma=00002333}).

The derivative for a deformation in the direction of the electric
field is
\begin{equation}
\left.\partial_{\xi_{1}}\epsilon_{r}\right|_{\xi=1}=-\left(\epsilon^{2}_{r}-\epsilon_{r}+2M\left(\epsilon_{r}-1\right)^{2}\right),\label{eq:Deriv1}
\end{equation}
and with this value the formula (\ref{eq:sigma11_solid}) also reproduces
the exact vacuum stress tensor component (\ref{eq:sigma=00002311})
in the direction of the electric field.

\subsection{A two-dimensional model}

\label{subsec:Exact2d}
\begin{figure}
\centering{}\includegraphics[scale=0.5]{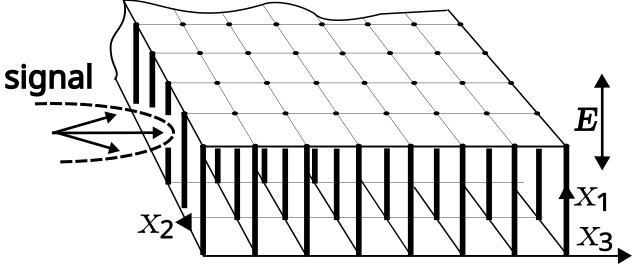}\caption{\label{fig:RodModel}A dielectric medium consisting regularly arranged
polarizable rods, all oriented in $x_{1}$-direction. The electric
field in the medium also points in $x_{1}$-direction.}
\end{figure}

Essential aspects of the problem can be illustrated
with an effectively two-dimensional model: a medium consisting of
thin polarizable rods arranged at the lattice points of a square lattice
in the $x_{1}$-$x_{3}$-plane (Fig.~\ref{fig:RodModel}). 
The electric field and all rods are oriented along the $x_{1}$-direction.
No polarization charge arises in the medium. The electric field thus
is constant, and macroscopic and microscopic electric fields are identical.
Only the component $\epsilon_{11}$ of the dielectric tensor plays
a role. If the system has length $\ell_{3}$ in $x_{3}$-direction,
then $\epsilon_{11}=\epsilon_{0}\left(1+\gamma_{a}/\ell_{3}\right)$,
where $\gamma_{a}$ is a constant. Eq.~(\ref{eq:sigma33_solid})
then again reproduces the (here trivial) vacuum stress tensor
\begin{equation}
\sigma_{22}=\sigma_{33}=\tfrac{1}{2}\left(\epsilon_{11}+\ell_{3}\tfrac{\partial\epsilon_{11}}{\partial\ell_{3}}\right)E^{2}_{1}=\tfrac{\epsilon_{0}}{2}E^{2}_{1}\label{eq:sigma11_rod}
\end{equation}
in the planes parallel to the rods. 

In the long wavelength limit, the model also permits exact solutions
for signal propagation in directions perpendicular to the $x_{1}$-axis,
for instance along $x_{3}$. Such signals only sense the dielectric
constant, not the individual rods. The dispersion relation is $\omega=ck\sqrt{\epsilon_{0}/\epsilon_{11}}.$
Adding the contribution from the magnetic field $B_{2}=-\sqrt{\epsilon_{r}}E_{1}$
according to (\ref{eq:sigma=000023def}) yields (all 1/2 time-average
factors are suppressed)
\begin{align*}
\sigma^{\mathrm{sign}}_{22}\left(E_{1},B_{2}\right) & =\tfrac{\epsilon_{0}}{2}\left(1-\epsilon_{r}\right)E^{2}_{1},\\
\sigma^{\mathrm{sign}}_{33}\left(E_{1},B_{2}\right) & =\tfrac{\epsilon_{0}}{2}\left(1+\epsilon_{r}\right)E^{2}_{1}.
\end{align*}
The boundaries of a signal with a finite width in $x_{2}$-direction
are pulled inward because of $\sigma^{\mathrm{sign}}_{22}<0$. Strictly
speaking, a finite width signal no longer represents purely two-dimensional
electrodynamics; electric and magnetic field cannot terminate abruptly,
and there is a transition region with oblique components of the fields.
However, the difference of the stress tensor inside and outside the
signal remains unchanged. If the rods form a two-dimensional liquid,
this inward force is compensated by a pressure. The effective longitudinal
stress tensor of a stationary signal thus is
\begin{equation}
\sigma^{\mathrm{sign,eff}}_{33}=\sigma^{\mathrm{sign}}_{33}-\sigma^{\mathrm{sign}}_{22}=\epsilon_{0}\epsilon_{r}E^{2}_{2}.\label{eq:sigma_33_sign2d}
\end{equation}
The corresponding momentum density is $\Pi_{3}=\sigma^{\mathrm{sign,eff}}_{33}\sqrt{\epsilon_{r}}/c$,
which gives
\begin{equation}
\tfrac{\mathrm{momentum}}{\mathrm{energy}}=\sqrt{\epsilon_{r}}/c.\label{eq:P_Minkowski}
\end{equation}
This is what is found in experiments in real liquids \cite{Jones_Rich_1954,Jones_Leslie_1978}.
Although the two-dimensional model is not directly applicable to real
liquids, it serves as an example where the underlying physics can
be understood starting from a microscopic framework.

\section{Arbitrarily oriented electric field}

We now generalize the computation of the stress tensor in a cubic
crystal in Section \ref{sec:sigma_cubic_E1} to a situation with an
arbitrarily oriented field $\boldsymbol{E}$. The electric field $\boldsymbol{E}_{\#}=\boldsymbol{E}+\tilde{\boldsymbol{E}}$
in the general case is the linear superposition of the field in the
situation with $\boldsymbol{E}$ along $x_{1}$ considered above,
with the field from the situation with $\boldsymbol{E}$ along $x_{2}$
or $x_{3}$. The field $\tilde{\boldsymbol{E}}$ as above comprises
the non-constant Fourier components of $\boldsymbol{E}_{\#}$.

It suffices to consider the plane $A_{1}$ given by $x_{1}=a/2$,
the field in the planes $A_{2}$ and $A_{3}$ follows by a cyclic
permutation of the coordinates. To simplify the notation we write
$qx=q_{2}x_{2}+q_{3}x_{3}$. The field component $E_{1}$ perpendicular
to $A_{1}$ considered above in (\ref{eq:ETildeTop}) yields the first
line of
\begin{align}
\tilde{E}_{1}\left(A_{1}\right) & =\left(\epsilon_{r}-1\right)E_{1}\left(\underline{1}+\sum\nolimits_{q\neq0}e^{iqx}\tfrac{aQ/2}{\sinh\tfrac{aQ}{2}}\right),\label{eq:ETildeA1}\\
\tilde{E}_{2}\left(A_{1}\right) & =-\left(\epsilon_{r}-1\right)\left(E_{2}\sum\nolimits_{q_{2}\neq0,q_{3}}e^{iqx}\tfrac{aq^{2}_{2}/4}{\tfrac{Q}{2}\sinh\tfrac{aQ}{2}}\right.\nonumber \\
 & \qquad\qquad+\left.E_{3}\sum\nolimits_{q_{2},q_{3}\neq0}e^{iqx}\tfrac{aq_{2}q_{3}/4}{\tfrac{Q}{2}\sinh\tfrac{aQ}{2}}\right),\nonumber \\
\tilde{E}_{3}\left(A_{1}\right) & =-\left(\epsilon_{r}-1\right)\left(E_{3}\sum\nolimits_{q_{2},q_{3}\neq0}e^{iqx}\tfrac{aq^{2}_{3}/4}{\tfrac{Q}{2}\sinh\tfrac{aQ}{2}}\right.\nonumber \\
 & \qquad\qquad\left.+E_{2}\sum\nolimits_{q_{2}\neq0,q_{3}}e^{iqx}\tfrac{aq_{2}q_{3}/4}{\tfrac{Q}{2}\sinh\tfrac{aQ}{2}}\right).\nonumber 
\end{align}
The field components $E_{2}$ and $E_{3}$ are parallel to $A_{1}$,
exactly as $E_{1}$ is parallel to $A_{3}$ in Eq.~(\ref{eq:ETildeSides}),
and analogously contribute diagonally as $\tilde{E}_{2}\propto E_{2}$
and $\tilde{E}_{3}\propto E_{3}$, and in the form $\tilde{E}_{2}\propto E_{3}$
and $\tilde{E}_{3}\propto E_{2}$. This gives the second and third
line in Eq.~(\ref{eq:ETildeA1}).

The requisite averages $\left\langle \tilde{E}_{i}\tilde{E}_{j}\right\rangle =a^{-2}\int_{a^{2}}\mathrm{d}x_{2}\mathrm{d}x_{3}E_{i}E_{j}$
according to Appendix~\ref{subsec:App_FT_avg}, are again given by
the sum of the products of the Fourier components. One must only keep
in mind that the constant in $\tilde{E}_{1}\left(A_{1}\right)$ combines
with $E_{1}$ to $\epsilon_{r}E_{1}$. The result is
\begin{align}
\left\langle \tilde{E}_{1}\tilde{E}_{i}\right\rangle ' & =\left(\epsilon_{r}-1\right)^{2}\left(E^{2}_{1}2M,\,-E_{1}E_{2}M,\,-E_{1}E_{3}M\right),\label{eq:EiEj_Avg}\\
\left\langle \tilde{E}_{2}\tilde{E}_{2}\right\rangle  & =\left(\epsilon_{r}-1\right)^{2}\left[E^{2}_{2}K+E^{2}_{3}\left(M-K\right)\right],\nonumber \\
\left\langle \tilde{E}_{3}\tilde{E}_{3}\right\rangle  & =\left(\epsilon_{r}-1\right)^{2}\left[E^{2}_{3}K+E^{2}_{2}\left(M-K\right)\right],\nonumber \\
K & =\left(\epsilon_{r}-1\right)^{2}\tfrac{1}{16}\sum\nolimits_{q_{2}\neq0,q_{3}}\tfrac{q^{4}_{2}}{\left(\tfrac{Q}{2}\sinh\tfrac{Q}{2}\right)^{2}}\cong0.160466.\nonumber 
\end{align}
The constant $M$ is defined in Eq.~(\ref{eq:M_Def}), the constant
$K$ is new. Using Eq.~(\ref{eq:sigma=000023def}) one can now compute
$\left\langle \tilde{\sigma}_{i1}\left(A_{1}\right)\right\rangle $.
Adding the contribution from the constant field $\left(\epsilon_{r}E_{1},E_{2},E_{3}\right)$
yields
\begin{align}
\sigma^{\mathrm{cubic}}_{11} & =\tfrac{\epsilon_{0}}{2}\left[E^{2}_{2}+E^{2}_{3}-\epsilon^{2}_{r}E^{2}_{1}+\left(\epsilon_{r}-1\right)^{2}M\left(E^{2}_{2}+E^{2}_{3}-2E^{2}_{1}\right)\right],\label{eq:sigma_i1}\\
\sigma^{\mathrm{cubic}}_{21} & =\tfrac{\epsilon_{0}}{2}E_{2}E_{1}\left[-2\epsilon_{r}+\left(\epsilon_{r}-1\right)^{2}2M\right],\nonumber \\
\sigma^{\mathrm{cubic}}_{31} & =\tfrac{\epsilon_{0}}{2}E_{3}E_{1}\left[-2\epsilon_{r}+\left(\epsilon_{r}-1\right)^{2}2M\right].\nonumber 
\end{align}
The constant $K$ has dropped out. The expressions (\ref{eq:sigma_i1})
and its cyclic permutations of the coordinates can be combined to
\begin{align}
\sigma^{\mathrm{cubic}}_{ij} & =\tfrac{\epsilon_{0}}{2}\left\{ \delta_{ij}\left[\boldsymbol{E}^{2}-\left(\epsilon_{r}-1\right)^{2}E^{2}_{i}\right]-2\epsilon_{r}E_{i}E_{j}\right\} \label{eq:sigma^cubic}\\
 & \qquad\qquad+\tfrac{\epsilon_{0}}{2}\left(\epsilon_{r}-1\right)^{2}M\left[\delta_{ij}\left(\boldsymbol{E}^{2}-5E^{2}_{i}\right)+2E_{i}E_{j}\right].\nonumber 
\end{align}
This is our final result for a cubic crystal. Omitting the areas
$A_{i}$ is justified since $\sigma$ is independent of the specific
areas in a translationally invariant system. Areas containing dipoles
can be slightly deformed to avoid possible complications inside the
dipoles. The vacuum stress tensor (\ref{eq:sigma^cubic}) evaluated
over interstitial surfaces $a^{2}$ enclosing some atoms determines
the electromagnetic force acting on them.

The stress tensor (\ref{eq:sigma^cubic}) contains the dielectric
tensor $\epsilon_{r}$ and the constant $M$ from (\ref{eq:M_Def}).
For $E_{i}=E_{1}\delta_{i,1}$ it reproduces Eq.~(\ref{eq:sigma=00002333})
and Eq.~(\ref{eq:sigma=00002311}). It strongly depends on the direction
of the field. One can now, for instance, compute $\sigma^{\mathrm{cubic}}_{ij}$
for an electric field $\boldsymbol{E}=E\hat{\boldsymbol{e}}$ along
a space diagonal $\hat{\boldsymbol{e}}$, and then project onto $\hat{\boldsymbol{e}}$
to get the electromagnetic stress $\sum\hat{e}_{i}\sigma^{\mathrm{cubic}}_{ij}\hat{e}_{j}$
in the direction of the field.

\section{Amorphous solids}

\label{sec:Amorph}A simple model of an amorphous solid is a mosaic
of randomly oriented grains of a cubic crystal. The dielectric tensor
of the mosaic is diagonal and agrees with that of the cubic crystal.
The electromagnetic stress tensor within the grains, however, is obtained
by averaging the stress tensor (\ref{eq:sigma^cubic}) over the orientations
of the grains. The electromagnetic stress tensor in the grain boundaries
is complicated, but the field $\boldsymbol{E}$ is finite, and the
contributions to the average stress tensor are negligibly when the
grains are large. 

One must distinguish between the stress tensor component in the direction
of the electric field and the component perpendicular to it. Projecting
Eq.~(\ref{eq:sigma^cubic}) onto the direction of $\boldsymbol{E}$
and averaging over all directions of $\boldsymbol{E}$ yields an expression
for the longitudinal electromagnetic stress tensor in an amorphous
solid,
\begin{align}
\sigma^{\mathrm{amorph}}_{\Vert}\left(\boldsymbol{E}\right) & =\sum_{ij}\left\langle \hat{E}_{i}\sigma^{\mathrm{cubic}}_{ij}\left(\boldsymbol{E}\right)\hat{E}_{j}\right\rangle _{\Omega}\label{eq:sigma^amorph||}\\
 & =\tfrac{\epsilon_{0}}{2}\boldsymbol{E}^{2}\left\langle \sum_{i}\left[\hat{E}^{2}_{i}-\left(\epsilon_{r}-1\right)^{2}\hat{E}^{4}_{i}\right]-2\epsilon_{r}\right.\nonumber \\
 & \qquad+\left.\left(\epsilon_{r}-1\right)^{2}M\left[\sum_{i}\hat{E}^{2}_{i}\left(1-5\hat{E}^{2}_{i}\right)+2\right]\right\rangle _{\Omega}\nonumber \\
 & =-\tfrac{\epsilon_{0}}{2}\boldsymbol{E}^{2}\left(2\epsilon_{r}-1+\tfrac{3}{5}\left(\epsilon_{r}-1\right)^{2}\right).\nonumber 
\end{align}
The angular average is denoted by $\left\langle \ldots\right\rangle _{\Omega}$.
To obtain $\sigma^{\mathrm{amorph}}_{\Vert}$ we have inserted Eq.~(\ref{eq:sigma^cubic})
and written $E_{i}=\left|\boldsymbol{E}\right|\hat{E}_{i}$. The result
then follows from
\begin{equation}
\left\langle \hat{E}^{2}_{i}\right\rangle {}_{\Omega}=\tfrac{1}{3},\qquad\left\langle \hat{E}^{4}_{i}\right\rangle {}_{\Omega}=\tfrac{1}{5},\label{eq:Avg_Omega_Basic}
\end{equation}
see Appendix~\ref{subsec:App_OmegaAvg}. The expression~(\ref{eq:sigma^amorph||})
differs from the longitudinal stress tensor (\ref{eq:sigma=00002311})
along an axis of a cubic crystal at order $\left(\epsilon_{r}-1\right)^{2}$.

To obtain the transverse electromagnetic stress tensor, one must project
Eq.~(\ref{eq:sigma^cubic}) onto a direction $\boldsymbol{k}$ perpendicular
to a given $\boldsymbol{E}$, average over $\boldsymbol{k}$, and
then over the directions of $\boldsymbol{E}$, 
\begin{align}
\sigma^{\mathrm{amorph}}_{\bot}\left(\boldsymbol{E}\right) & =\sum_{ij}\left\langle k_{i}\sigma^{\mathrm{cubic}}_{ij}\left(\boldsymbol{E}\right)k_{j}\right\rangle _{\Omega}\label{eq:sigma^amorph|_}\\
 & =\tfrac{\epsilon_{0}}{2}\boldsymbol{E}^{2}\sum_{i}\left\langle k^{2}_{i}\left[1-\left(\epsilon_{r}-1\right)^{2}\hat{E}^{2}_{i}\right]\right.\nonumber \\
 & \qquad+\left.\left(\epsilon_{r}-1\right)^{2}Mk^{2}_{i}\left(1-5\hat{E}^{2}_{i}\right)\right\rangle _{\Omega}\nonumber \\
 & =\tfrac{\epsilon_{0}}{2}\boldsymbol{E}^{2}\left(1-\tfrac{1}{5}\left(\epsilon_{r}-1\right)^{2}\right).\nonumber 
\end{align}
The average over $\boldsymbol{k}$ in $\left\langle E^{2}_{i}k^{2}_{i}\right\rangle _{\Omega}$
can be written as an average over an angle $\varphi$ by expressing
$\boldsymbol{k}$ as a sum of two unit vectors orthogonal to $\boldsymbol{E}$,
\begin{align*}
\boldsymbol{k} & =\boldsymbol{u}\sin\varphi+\boldsymbol{v}\cos\varphi,\\
\boldsymbol{u} & =\left(E_{2},-E_{1},0\right)/\sqrt{E^{2}_{1}+E^{2}_{2}},\\
\boldsymbol{v} & =\hat{\boldsymbol{E}}\times\boldsymbol{u}.
\end{align*}
This gives $\left\langle k^{2}_{i}\right\rangle _{\varphi}=\tfrac{1}{2}\left(u^{2}_{i}+v^{2}_{i}\right)=\tfrac{1}{2}\left(1-\hat{E}^{2}_{i}\right)$.
The last equal sign simply states that the squares of the components
in each row of the orthogonal matrix $\left(\boldsymbol{u},\boldsymbol{v},\hat{\boldsymbol{E}}\right)$
sum to one. The last line in (\ref{eq:sigma^amorph|_}) then follows
from
\begin{align*}
\left\langle \hat{E}^{2}_{i}k^{2}_{i}\right\rangle _{\Omega} & =\tfrac{1}{2}\left\langle \hat{E}^{2}_{i}\left(1-\hat{E}^{2}_{i}\right)\right\rangle _{\Omega}=1/15,
\end{align*}
see also Eq.~(\ref{eq:Avg_Omega_Basic}). The stress tensor components
(\ref{eq:sigma^amorph||}) and (\ref{eq:sigma^amorph|_}) in an amorphous
solid no longer depend on the constant $M$. This, however, does not
mean that they can be derived from macroscopic electrodynamics alone.
The transverse component~(\ref{eq:sigma^amorph|_}), notably, agrees
with Peierls's result for amorphous solids and liquids~\cite{Peierls1976}. 

\section{Conclusion}

Idealized models with point dipoles on a lattice are usually quantitatively
useful for real crystals as well. We have computed the electric field
in an orthorhombic lattice of point dipoles, first with the macroscopic
electric field oriented along a crystal axis. This has allowed us
to derive the electromagnetic stress tensor in a cubic crystal in
two ways: directly from the microscopic field, and alternatively,
using the conventional phenomenological formula, involving the macroscopic
electric field and a derivative of the dielectric tensor. The results
agree. The fact that the stress tensor contains a constant $M$ depending
on the microscopic details, rules out its derivation from macroscopic
electrodynamics alone~\cite{Shevch_2010,Frias_2012}, at least for
cubic crystals.

The generalization to an arbitrarily oriented macroscopic electric
field in a cubic crystal is essentially a linear superposition of
the more symmetric case. The generic electromagnetic stress tensor
$\sigma^{\mathrm{cubic}}_{ij}$ is still defined by its value in the
interstitial vacuum planes. 

Modeling an amorphous solid as a mosaic of arbitrarily oriented grains
of a cubic crystal has allowed to compute the longitudinal and transverse
components of $\sigma$ in an amorphous solid as an average of $\sigma^{\mathrm{cubic}}_{ij}$
over crystal orientations. The constant $M$ drops out. 

The propagation of light in a liquid represents a physical system
where the electromagnetic stress tensor can be directly measured,
making it the subject of theoretical investigations. Although dynamic
phenomena are generally beyond the scope of this work\textendash apart
from the two-dimensional model in Section \ref{subsec:Exact2d} \textendash several
concluding remarks are in order.

The experiments~\cite{Jones_Rich_1954,Jones_Leslie_1978} utilized
a signal of finite width, and the measured signal momentum is given
by the ratio~(\ref{eq:P_Minkowski}). 

In his analysis, Peierls~\cite{Peierls1976} started from a partially
microscopic model and also arrived at the stress tensor component
(\ref{eq:sigma^amorph|_}) in propagation direction. The corresponding
usual magnetic contribution directly follows from the macroscopic
Maxwell equations. 

What then would be needed to get the ratio (\ref{eq:P_Minkowski})
is a mechanism eliminating the unusual factor of $1/5$ from the electric
contribution (\ref{eq:sigma^amorph|_}). The two-dimensional model
discussed in Section~\ref{subsec:Exact2d} indicates at least that
the finite width of the signal\textendash and the stress tensor components
in the directions of $\boldsymbol{E}$ and $\boldsymbol{B}$ at the
lateral borders of the signal\textendash play a role. 

\bigskip{}

\bibliographystyle{habbrv}
\bibliography{Other}

\appendix

\section{Appendix}

\subsection{Fourier sums}

\label{subsec:App_FT_sum}To compute the electric field in interstitial
planes, the following equation was used: 
\begin{equation}
f_{Q}\left(z\right)=\sum_{q\in2\pi\mathbb{Z}/a}\tfrac{e^{iqz}}{Q^{2}+q^{2}}=\tfrac{a^{2}}{4}\tfrac{\cosh\left(Q\left(z-\tfrac{a}{2}\right)\right)}{\tfrac{Qa}{2}\sinh\left(\tfrac{Qa}{2}\right)}.\label{eq:MatsubaraSum}
\end{equation}
This equation is valid in the interval $0\leq z\leq a$ for any $Q\neq0$,
and must be extended periodically such that $f_{Q}\left(z\right)=f_{Q}\left(z+a\right)$.
To verify the Fourier sum, it suffices to check that both expressions
satisfy the differential equation 
\[
\left(\partial^{2}_{z}-Q^{2}\right)f_{Q}\left(z\right)=-a\sum_{m\in\mathbb{Z}}\delta\left(z-ma\right).
\]
The r.h.s. satisfies the equation for $z\notin\mathbb{Z}a$ since
it is a linear combination of $e^{\pm Qz}$. The inhomogeneous differential
equation is satisfied because $f_{Q}'\left(0^{+}\right)-f_{Q}'\left(0^{-}\right)=2f_{Q}'\left(0^{+}\right)=-a$.
This discontinuous derivative at $z=0$ generates a $\delta$-function
in the second derivative. One also needs the Fourier series
\begin{equation}
g\left(z\right)=\sum_{q\in2\pi\mathbb{Z}/a\setminus0}\tfrac{e^{iqz}}{q^{2}}=\tfrac{1}{2}\left(z^{2}-za+\tfrac{1}{6}a^{2}\right)=\lim_{Q\rightarrow0}\left(f_{Q}\left(z\right)-1/Q^{2}\right),\label{eq:MatsubaraSum2}
\end{equation}
valid in the interval $0\leq z\leq a$. Instead of deriving $g\left(z\right)$
as a limiting case of $f_{Q}\left(z\right)$ one could alternatively
verify that $g\left(z\right)$ satisfies the differential equation
$g''\left(z\right)=1-a\sum_{m\in\mathbb{Z}}\delta\left(z-ma\right)$.

\subsection{Average over a surface of a unit cell}

\label{subsec:App_FT_avg}The average stress tensor (\ref{eq:sigma=000023def})
in the plane $x_{3}=a/2$ of a cubic crystal contains averages of
the type
\[
\left\langle s_{i,j}\right\rangle =a^{-2}\int^{a}_{0}\mathrm{d}x\int^{a}_{0}\mathrm{d}yE^{\#}_{i}\left(x,y\right)E^{\#}_{j}\left(x,y\right).
\]
The components of the electric field in the area are given as a Fourier
series
\begin{equation}
E^{\#}_{i}\left(x,y\right)=\sum_{\left\{ p,q\right\} \epsilon\left(2\pi\mathbb{Z}/a\right)^{2}}e_{i}\left(p,q\right)e^{ipx+qy},\label{eq:E_i(xy)}
\end{equation}
as in (\ref{eq:ETildeSides}). Since $E^{\#}_{i}\left(x,y\right)$
is real it follows $e^{*}_{i}\left(p,q\right)=e_{i}\left(-p,-q\right)$.
Inserting the sum (\ref{eq:E_i(xy)}) yields
\begin{align*}
\left\langle s_{i,j}\right\rangle  & =a^{-2}\int\mathrm{d}x\mathrm{d}y\sum_{p,q,p',q'}e_{i}\left(p,q\right)e^{ipx+qy}e_{j}\left(p',q'\right)e^{ip'x+q'y}\\
 & =\sum_{p,q}e_{i}\left(p,q\right)e_{j}\left(-p,-q\right)=\sum_{\left\{ p,q\right\} \epsilon\left(2\pi\mathbb{Z}/a\right)^{2}}e_{i}\left(p,q\right)e^{*}_{j}\left(p,q\right).
\end{align*}
The field $E_{i}$ in $E^{\#}_{i}\left(x,y\right)=E_{i}+\tilde{E}_{i}\left(x,y\right)$
only contributes as $e_{i}\left(0,0\right)=E_{i}+\tilde{e}_{i}\left(0,0\right)$,
where $\tilde{e}_{i}\left(0,0\right)$ is the $p=q=0$ component of
$\tilde{E}_{i}\left(x,y\right)$.

\subsection{Energy balance in a solid}

\label{subsec:App_EnergyBalance}The phenomenological expression (\ref{eq:sigma_LL})
for the electromagnetic stress tensor of a liquid directly follows
from energy conservation~\cite{LL_FT1971}. Its generalization (\ref{eq:sigma11_solid},
\ref{eq:sigma33_solid}) to an orthorhombic crystal or an amorphous
solid is straightforward. Consider a planar capacitor with plates
of area $A$ perpendicular to the $x_{1}$-axis, separated by a distance
$\ell_{1}$, and an electric field $E_{1}$. At constant charge $Q$,
the total energy is
\[
W=\tfrac{\ell_{1}}{2\epsilon_{11}A}Q^{2}=\tfrac{\epsilon_{11}}{2}A\ell_{1}E^{2}_{1}.
\]
Fringe effects at the borders are negligible if the area $A$ is large.
Under a strictly uniaxial deformation, the variation of the energy
with distance $\ell_{1}$ yields the stress tensor component
\begin{align*}
\sigma^{\mathrm{solid}}_{11}\left(E_{1}\right) & =-\tfrac{1}{A}\left(\tfrac{\partial W}{\partial\ell_{1}}\right)_{Q}=-\tfrac{1}{2}\left(\tfrac{1}{\epsilon_{11}}-\tfrac{\ell_{1}}{\epsilon^{2}_{11}}\tfrac{\partial\epsilon_{11}}{\partial\ell_{1}}\right)\tfrac{Q^{2}}{A^{2}}\\
 & =-\tfrac{1}{2}\left(\epsilon_{11}-\ell_{1}\tfrac{\partial\epsilon_{11}}{\partial\ell_{1}}\right)E^{2}_{1}.
\end{align*}
It is clear from the derivation that the stress tensor does not include
the electromagnetic stress within the atoms, which is compensated
by some other force. It could be computed, however, from the microscopic
field in an interstitial vacuum plane, if such a plane exists. The
expression (\ref{eq:sigma33_solid}) for the stress tensor in the
direction perpendicular to the field is obtained by varying of $A=\ell_{2}\ell_{3}$.

\subsection{Average over orientations}

\label{subsec:App_OmegaAvg}In Section \ref{sec:Amorph} the averages
(\ref{eq:Avg_Omega_Basic}) of unit vector components over all orientations
were used to calculate the stress tensor of an amorphous solid from
that of a cubic crystal. These averages can be derived using spherical
coordinates and the unit vector component $\hat{z}=\cos\theta$,
\[
\left\langle \hat{z}^{2m}\right\rangle _{\Omega}=\tfrac{2\pi}{4\pi}\int^{\pi}_{0}\mathrm{d}\theta\sin\theta\cos^{2m}\theta=\tfrac{1}{2}\tfrac{-1}{2m+1}\cos^{2m+1}\theta|^{\pi}_{0}=\tfrac{1}{2m+1}.
\]

\vfill{}
\end{document}